\documentclass[12pt]{article}

\usepackage{graphicx}

\def\gtwid{\mathrel{\raise.3ex\hbox{$>$\kern-.75em\lower1ex\hbox{$\sim$}}}}
\def\ltwid{\mathrel{\raise.3ex\hbox{$<$\kern-.75em\lower1ex\hbox{$\sim$}}}}
\def\square{\kern1pt\vbox{\hrule height 1.2pt\hbox{\vrule width 1.2pt\hskip 3pt
   \vbox{\vskip 6pt}\hskip 3pt\vrule width 0.6pt}\hrule height 0.6pt}\kern1pt}

\begin{document}

\begin{titlepage}

\begin{flushright}
UFIFT-QG-16-09 \\
CCTP-2016-19 \\
CCQCN-2016-178
\end{flushright}

\vskip 2cm

\begin{center}
{\bf Causality Implies Inflationary Back-Reaction}
\end{center}

\vskip 1cm

\begin{center}
S. Basu$^{1*}$, N. C. Tsamis$^{2\star}$ and R. P. Woodard$^{1\dagger}$
\end{center}

\vskip .5cm

\begin{center}
\it{$^{1}$ Department of Physics, University of Florida,\\
Gainesville, FL 32611, UNITED STATES}
\end{center}

\begin{center}
\it{$^{2}$ Institute of Theoretical Physics \& Computational Physics, \\
Department of Physics, University of Crete, \\
GR-710 03 Heraklion, HELLAS}
\end{center}

\vspace{1cm}

\begin{center}
ABSTRACT
\end{center}
There is a widespread belief among inflationary cosmologists that
a local observer cannot sense super-horizon gravitons. The argument
goes that a local observer would subsume super-horizon gravitons
into a redefinition of his coordinate system. We show that adopting 
this view for pure gravity on de Sitter background leads to time 
variation in the Hubble parameter measured by a local observer. It 
also leads to a violation of the gravitational field equation $R = 
4 \Lambda$ because that equation is obeyed by the full metric,
rather than the one which has been cleansed of super-horizon modes.

\begin{flushleft}
PACS numbers: 04.50.Kd, 95.35.+d, 98.62.-g
\end{flushleft}

\vskip .5cm

\begin{flushleft}
$^{*}$ e-mail: shinjinibasu@ufl.edu \\
$^{\star}$ e-mail: tsamis@physics.uoc.gr \\
$^{\dagger}$ e-mail: woodard@phys.ufl.edu
\end{flushleft}

\end{titlepage}

\section{Introduction}

One of the peculiar features of quantum field theory during inflation
is the existence of secular corrections from loops of massless, minimally
coupled scalars \cite{Onemli:2002hr,Prokopec:2002uw,Prokopec:2003qd,
Brunier:2004sb,Miao:2006pn,Prokopec:2006ue,Prokopec:2008gw,Janssen:2009pb,
Kahya:2009sz,Kahya:2010xh,Kitamoto:2010et,Park:2011ww,Kitamoto:2011yx}
and/or gravitons \cite{Tsamis:1996qm,Tsamis:1996qk,Miao:2005am,
Kahya:2007bc,Kitamoto:2012ep,Kitamoto:2012vj,Miao:2012bj,Leonard:2013xsa,
Boran:2014xpa,Glavan:2015ura}. 
These corrections grow without bound as long as inflation lasts.
Among the many interesting effects caused by these secular corrections are 
changes to particle kinematics \cite{Prokopec:2002jn,Prokopec:2003iu,
Garbrecht:2006jm,Miao:2006gj,Kahya:2006hc,Kahya:2007cm,Mora:2013ypa,
Wang:2014tza, Glavan:2016bvp}, changes in long range forces 
\cite{Kitamoto:2012ep,Kitamoto:2012vj,Degueldre:2013hba,Glavan:2013jca,
Park:2015kua}, changes in the primordial power spectra 
\cite{Weinberg:2005vy,Weinberg:2006ac,Kahya:2010xh}, and changes in the 
cosmic expansion rate \cite{Tsamis:1996qq,Onemli:2004mb,Miao:2006pn,
Prokopec:2007ak}.

There is no dispute about the reality of secular corrections driven
by massless, minimally coupled scalars. The stochastic formalism of
Starobinsky \cite{Starobinsky:1986fx} even provides a method for working 
out what happens at late times in those cases which approach a static 
limit \cite{Starobinsky:1994bd,Woodard:2005cw,Tsamis:2005hd,Miao:2006pn,
Prokopec:2007ak,Finelli:2008zg,Finelli:2010sh}. On the other hand, 
there are fierce debates concerning both the reality of secular 
corrections from inflationary gravitons and what they might do after 
perturbation theory breaks down \cite{Garriga:2007zk,Tsamis:2008zz}.

Secular effects derive from more and more of the plane wave mode 
functions of free scalar and graviton fields approaching a nonzero, 
spacetime constant \cite{Vilenkin:1982wt,Linde:1982uu,Starobinsky:1982ee}.
On de Sitter background both mode functions are \footnote{The small
$k$ and late time limiting forms of $u(\eta,k)$ are the same, which
has led to much confusion between infrared divergences and secular 
growth. The former derive from the region near $k = 0$, while secular
growth derives from the region near $k = H a(\eta)$, which grows without 
bound.}
\begin{equation}
u(\eta,k) = \frac{H}{\sqrt{2 k^3}} \Bigl[1 - \frac{i k}{H a(\eta)} \Bigr] 
\exp\Bigl[\frac{i k}{H a(\eta)} \Bigr] \longrightarrow \frac{H}{\sqrt{2 k^3}} 
\quad , \quad a(\eta) = -\frac1{H \eta} \; . \label{modefunc} \end{equation}
Their approach to a constant is known as ``freezing in'' and it is how 
inflationary perturbations survive to much later times. Physicists accept 
that the freezing in of scalars can mediate effects because the value of a 
scalar field is observable; for example, the expectation value of the Higgs 
field determines masses in the Standard Model. However, physicists are 
conditioned to dismiss nonzero constant values of the graviton field as 
gauge artifacts which could be eliminated by an appropriate choice of 
coordinates \cite{Higuchi:2011vw,Miao:2011ng,Morrison:2013rqa,Miao:2013isa,
Frob:2014fqa,Woodard:2015kqa}.

This belief is problematic because the graviton mode functions are not
exactly constant. They vary rapidly at early times ($k \gg H a$), and it 
was never clear why their passage to the late time regime ($k \ll H a$) 
can have no observable effect beyond the tensor power spectrum 
\cite{Miao:2012xc}. However, a supporting argument is adduced based on 
the presumed difficulty of a local observer in resolving the spacetime 
variation of any mode whose wavelength exceeds the causal horizon 
($k < H a(\eta)$) \cite{Urakawa:2009my,Urakawa:2009gb}. It is asserted 
that a local observer would instead subsume these super-horizon modes 
into a transformation of his coordinate system \cite{Giddings:2010nc,
Urakawa:2010it,Tanaka:2011aj,Giddings:2011zd}. We refer to this belief 
as the {\it Transformation Ansatz} \cite{Basu:2016iua}.

There are several reasons to doubt the Transformation Ansatz. First, field 
theory interactions are local in spacetime, not in Fourier space. 
Variations of long wavelength modes are indeed difficult to resolve, 
but this is a consequence of local dynamics and does not require the 
excision of dynamical variables by some {\it deux ex machina}. Second, 
the range of excluded $k$ values changes with time: it increases during 
inflation and decreases after the end of inflation. Finally, the fact 
that inflation ends means that the Hubble radius is not a true horizon, 
nor is there any invariant meaning to wave number $k$ in the full, 
interacting theory.

Yet the Transformation Ansatz has many adherents \cite{Senatore:2012nq,
Senatore:2012wy,Pimentel:2012tw,Tanaka:2012wi,Tanaka:2013xe,
Tanaka:2014ina,Tanaka:2015aza}, and it is invoked to 
deny the possibility of secular graviton corrections in general,
and of secular back-reaction in particular. Our purpose here is not to 
pass on the validity of the Transformation Ansatz but rather to demonstrate 
that {\it adopting it leads to precisely the opposite conclusion about 
secular back-reaction}. The reason should become clear when we carefully 
examine what happens to the conformal factor upon attempting 
to absorb super-horizon gravitons into a redefinition of coordinates. If 
$h_{\mu\nu}(x)$ represents the graviton field that is approaching a constant, 
the actual metric is not $\eta_{\mu\nu} + \kappa h_{\mu\nu}(x)$ --- which 
really would be trivial if constant --- but rather,
\begin{equation}
g_{\mu\nu}(x) = a^2 \Bigl[ \eta_{\mu\nu} + \kappa h_{\mu\nu}\Bigr] \equiv
a^2 \widetilde{g}_{\mu\nu} \qquad \kappa^2 \equiv 16 \pi G \; .
\end{equation}
The coordinate transformation which carries $\widetilde{g}_{\mu\nu}$ to
$\eta_{\mu\nu}$ --- under the false assumption that $\widetilde{g}_{\mu\nu}$
is exactly constant --- changes how the scale factor depends upon the new
time coordinate. This leads to a secular decrease of the Hubble parameter
and a violation of the field equation.

In section 2 we give the transformation as a function of the graviton field,
assuming (according to the Transformation Ansatz) that it is exactly constant. 
In section 3 we show that the expected Hubble parameter decreases at order 
$\kappa^4$, and that there is a corresponding violation of the Einstein 
equation. Section 4 discusses the fascinating question of what this all might 
mean.

\section{The Transformation}

We deal with three different metrics:
\begin{itemize}
\item{The true metric $g_{\mu\nu}(x)$, which includes the scale factor and
super-horizon modes;}
\item{The conformally rescaled metric $\widetilde{g}_{\mu\nu}(x) \equiv
g_{\mu\nu}(x)/a^2$ which has the conformal factors cancelled but still
 contains super-horizon modes; and}
\item{The local observer's metric $\mathbf{g}_{\mu\nu}(x)$ which has been
cleansed of super-horizon modes by subsuming them into a coordinate
redefinition.}
\end{itemize}
We also decompose the conformally rescaled metric $\widetilde{g}_{\mu\nu}(x)$
into a super-horizon part $\gamma_{\mu\nu}(x) = \eta_{\mu\nu} + \kappa
\psi_{\mu\nu}(x)$, and a sub-horizon part $\kappa \chi_{\mu\nu}(x)$,
\begin{equation}
\widetilde{g}_{\mu\nu}(x) \equiv \gamma_{\mu\nu}(x) + \kappa \chi_{\mu\nu}(x)
\; .
\end{equation}

The local observer's metric is defined by constructing the linear coordinate
transformation $x^{\mu} \rightarrow {x'}^{\mu}$ which would carry
$\gamma_{\mu\nu}(x)$ to $\eta_{\mu\nu}$ under the Transformation Ansatz 
assumption that $\gamma_{\mu\nu}(x)$ is exactly constant. We use the matrix 
coefficients $\omega^{\mu}_{~\nu}$ to denote the inverse transformation 
$x^{\mu} = \omega^{\mu}_{~\nu} {x'}^{\nu}$. The local observer's metric is,
\begin{equation}
\mathbf{g}_{\mu\nu}(x) \equiv \omega^{\rho}_{~\mu} \omega^{\sigma}_{~\nu}
g_{\rho\sigma}(\omega x) = a^2\Bigl(\omega^0_{~\alpha} x^{\alpha}\Bigr)
\times \Bigl[\eta_{\mu\nu} + \omega^{\rho}_{~\mu} \omega^{\sigma}_{~\nu}
\kappa \chi_{\rho\sigma}(\omega x) \Bigr] \; . \label{localmetric}
\end{equation}

In addition to having the property $\omega^{\rho}_{~\mu} \omega^{\sigma}_{~\nu}
\gamma_{\mu\nu}(x) = \eta_{\mu\nu}$ we want the local observer's scale factor to
depend only on conformal time, which means $\omega^0_{~i} = 0$. The solution
for $\omega^{\mu}_{~\nu}$ turns out to be the Lorentz-symmetric vierbein
\cite{Woodard:1984sj} with a Lorentz boost to null the time-space components
\cite{Basu:2016iua},
\begin{equation}
\omega^{\mu}_{~\nu} \equiv \left( \matrix{ \omega^0_{~0} & \omega^0_{~n} \cr
\cr \omega^{m}_{~0} & \omega^{m}_{~n}} \right) =
\left( \matrix{ \frac1{N} & 0 \cr \cr \frac{N^m}{N} & e^{m}_{~n} }\right) \; ,
\label{trans1}
\end{equation}
where $N$ and $N^m$ are the lapse and shift \cite{Deser:1959zza,
Arnowitt:1960es,Arnowitt:1962hi} of $\gamma_{\mu\nu}$, and
$e^{m}_{~n}$ is the inverse driebein of its spatial components
$\gamma_{mn} = \Gamma_{mn}$ (i.e., $\Gamma^{mn} = e^{m}_{~k} e^{n}_{~k}$),
\begin{equation}
\frac1{N} = \sqrt{-\gamma^{00}} \quad , \quad
\frac{N^m}{N} = -\frac{\gamma^{0 m}}{\sqrt{-\gamma^{0 0}}} \quad , \quad
e^{m}_{~n} = \Bigl(\sqrt{\Gamma^{-1} \times I} \, \Bigr)^{m}_{~n} \; .
\label{trans2}
\end{equation}
The transformation (\ref{trans1}-\ref{trans2}) is unique up to a
3-rotation of the inverse driebein, $e^{m}_{~n} \rightarrow e^{m}_{~k}
\times R_{k n}$, which plays no role for us.

We should emphasize that $x^{\mu} = \omega^{\mu}_{~ \nu} {x'}^{\nu}$ is
not a true coordinate transformation because the matrix 
$\omega^{\mu}_{~\nu}$ given by (\ref{trans1}-\ref{trans2}) is not
a spacetime constant,
\begin{equation}
\partial_{\rho} \omega^{\mu}_{~\sigma} \neq \partial_{\sigma}
\omega^{\mu}_{~\rho} \qquad \Longrightarrow \qquad \omega^{\mu}_{~\nu}
\neq \frac{\partial x^{\mu}}{\partial {x'}^{\nu}} \; . \label{obstacle}
\end{equation}
The principle obstacle in (\ref{obstacle}) is not the small residual 
spacetime dependence of any particular super-horizon mode in 
$\psi_{\mu\nu}(x)$ but rather the fact that more and more modes make 
the transition from $\chi_{\mu\nu}(x)$ to $\psi_{\mu\nu}(x)$ as they 
experience horizon crossing. This means we must view the local
observer's metric (\ref{localmetric}) as a nonlocal field redefinition 
of the original metric, which may not be expanding at the same rate
and may not even obey the same local field equation.

\section{Back-Reaction}

Expressions (\ref{localmetric}) and (\ref{trans1}-\ref{trans2}) imply
that the local observer's scale factor is,
\begin{equation}
a\Bigl( \omega^0_{~\rho} x^{\rho}\Bigr) = -\frac1{H \sqrt{-\gamma^{00}}
\times \eta} \; . \label{localscale}
\end{equation}
This is de Sitter with a rescaled Hubble constant which depends slightly
on spacetime through the super-horizon part of the graviton field,
\begin{equation}
\mathbf{H}(x) \equiv H \times \sqrt{-\gamma^{00}(x)} \; , \label{localH}
\end{equation}
The residual time dependence of super-horizon mode functions
is not significant but the continual addition of {\it new} modes --- 
as they pass from the sub-horizon graviton field $\chi_{\mu\nu}(x)$ 
to the super-horizon graviton field $\psi_{\mu\nu}(x)$
--- introduces an appreciable time dependence. We begin by expanding
the operator $\mathbf{H}(x)$ in powers of $\psi_{\mu\nu}$. We then
compute the expectation value at 4th order, and the section closes
with a discussion of the field equation obeyed by the local observer's
metric $\mathbf{g}_{\mu\nu}(x)$.

One might wonder about the effect of modes which are super-horizon even on the 
initial value surface. This is a fascinating question whose answer we do not
know. However, there is a simple way to distinguish these initially 
super-horizon modes from initially sub-horizon modes which experience first
horizon crossing during the course of inflation. This is just to work on a 
finite spatial manifold such as $T^3$, which supports spatially flat de Sitter
background. If the initial physical radius of the manifold is smaller than the 
Hubble length then there are no initially super-horizon modes. This is also a
standard technique for controlling infrared divergences \cite{Tsamis:1993ub}.

\subsection{The graviton expansion of $\mathbf{H}(x)$}

\begin{figure}[ht]
\begin{center}
\hspace{-2cm} \includegraphics[width=4cm,height=3cm]{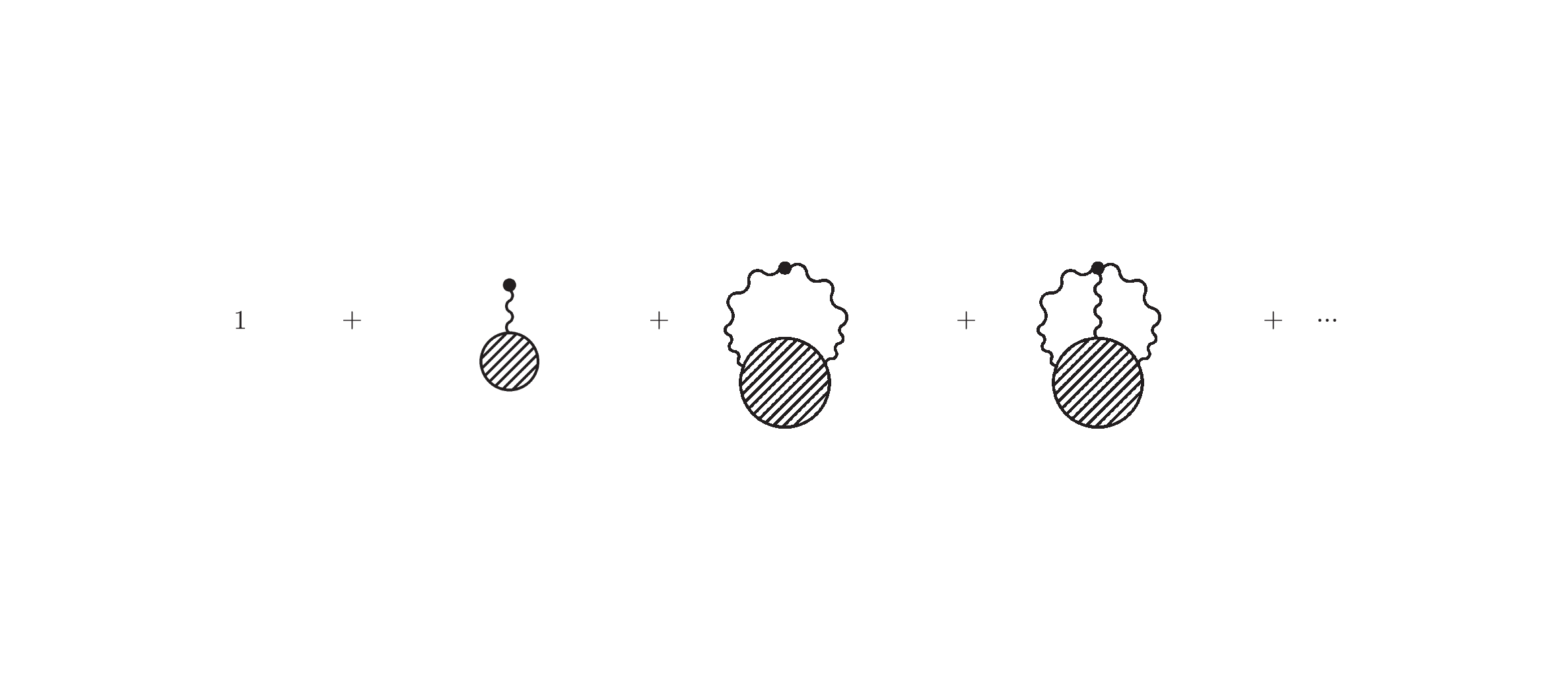}
\end{center}
\vskip -2.5cm
\caption{Expanding $\mbox{H}(x)/H$ in powers of the graviton field
and then taking the expectation value results in the diagrammatic
series depicted above. The spacetime point $x^{\mu}$ is the solid vertex.}
\label{Diagram1}
\end{figure}

Inverting $\gamma_{\mu\nu}(x) = \eta_{\mu\nu} + \kappa \psi_{\mu\nu}$,
employing the usual convention about raising and lowering graviton 
indices with $\eta_{\mu\nu}$, and taking account of the spacelike 
signature gives,
\begin{equation}
\gamma^{00} = -1 + \kappa \psi_{00} - \kappa^2 \psi_{0}^{~ \rho} 
\psi_{\rho 0} + \kappa^3 \psi_{0 \rho} \psi^{\rho}_{~ \sigma}
\psi^{\sigma}_{~ 0} - \kappa^4 \psi_{0 \rho} \psi^{\rho}_{~ \sigma}
\psi^{\sigma}_{~ \tau} \psi^{\tau}_{~ 0} + O(\kappa^5) \; .
\end{equation}
Substituting in expression (\ref{localH}) results in the expansion,
\begin{eqnarray}
\lefteqn{\frac{\mathbf{H}(x)}{H} = \frac1{N(x)} = 
\sqrt{-\gamma^{00}(x)} \; , } \\
& & \hspace{-.5cm} = 1 + \frac{\kappa}{2} \psi_{00} - \frac{\kappa^2}{2} 
\Bigl[ \psi_{0 \rho} \psi^{\rho}_{~ 0} + \frac14 \psi_{00}^2\Bigr] 
+ \frac{\kappa^3}{2} \Bigl[ \psi_{0\rho} \psi^{\rho}_{~ \sigma}
\psi^{\sigma}_{~ 0} + \frac12 \psi_{0 \rho} \psi^{\rho}_{~ 0} \psi_{00}
+ \frac1{8} \psi_{00}^3\Bigr] \nonumber \\
& & \hspace{.5cm} -\frac{\kappa^4}{2} \Bigl[\psi_{0\rho} 
\psi^{\rho}_{~ \sigma} \psi^{\sigma}_{~\tau} \psi^{\tau}_{~ 0} + 
\frac14 (\psi_{0\rho} \psi^{\rho}_{~0} )^2 + \frac12 \psi_{0\rho} 
\psi^{\rho}_{~ \sigma} \psi^{\sigma}_{~ 0} \psi_{00} \nonumber \\
& & \hspace{6cm} + \frac{3}{8} \psi_{0\rho} \psi^{\rho}_{~0}
\psi_{00}^2 + \frac{5}{64} \psi_{00}^4 \Bigr] + O(\kappa^5) \; . \qquad
\label{Hexpand}
\end{eqnarray}

\vskip .8cm

\begin{figure}[ht]
\begin{center}
\hspace{-3cm} \includegraphics[width=4cm,height=3cm]{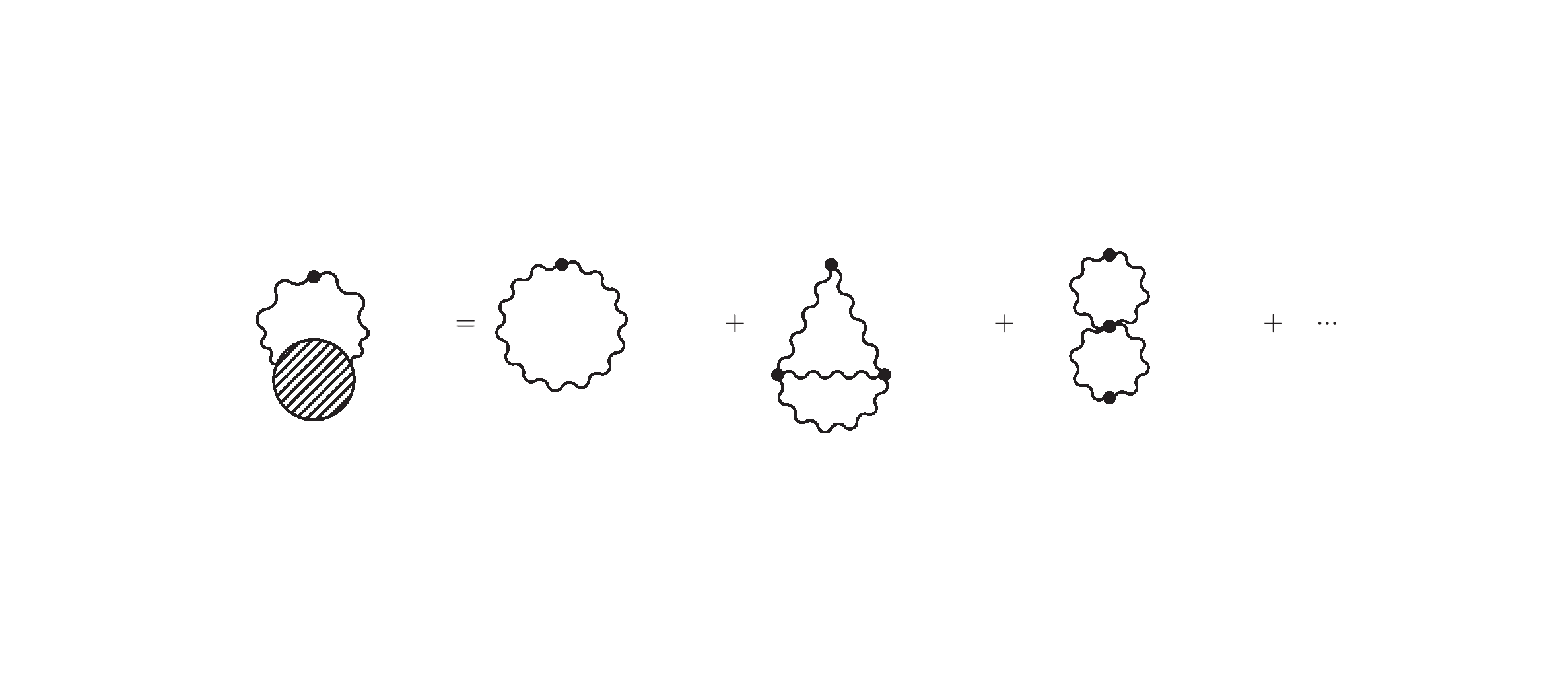}
\end{center}
\vskip -2.5cm
\caption{Each of the $N$-point functions of Fig.~\ref{Diagram1} 
has a conventional diagrammatic expansion. This graph shows the
lowest terms which contribute to the 2-point function.}
\label{Diagram2}
\end{figure}

Expression (\ref{Hexpand}) is a series of quantum operators. Even 
after horizon crossing these operators are superpositions of random
numbers, so we can only discuss the statistical properties of 
$\mathbf{H}(x)$ rather than its numerical value. Taking the expectation
value gives a series of diagrams having the general form shown in 
Figure~\ref{Diagram1}. Basically, any term in expression (\ref{Hexpand}) 
which contains $N$ factors of the super-horizon graviton field 
$\psi_{\mu\nu}(x)$ corresponds to a diagram with $N$ lines emanating
from the top vertex and then joining with all possible interactions.
For example, Figure~\ref{Diagram2} shows the expansion for $N=2$.

To understand the diagrams in Figures~\ref{Diagram1} and \ref{Diagram2} 
quantitatively it is important to go beyond linearized order in the free 
field expansion, and to be precise about what we mean by the sub-horizon 
and super-horizon parts. The Heisenberg field equations of general 
relativity permit us to express the full graviton field $h_{\mu\nu}(x)$ 
as a series in powers of the linearized solutions $h_{\mu\nu}^{(1)}(x)$ 
about de Sitter background,
\begin{equation}
h_{\mu\nu}(x) = h_{\mu\nu}^{(1)}(x) + \kappa h_{\mu\nu}^{(2)}(x) +
\kappa^2 h_{\mu\nu}^{(3)}(x) + \dots 
\end{equation}
Here $h_{\mu\nu}^{(m)}$ represents the term in the full solution
which contains $m$ factors of the free field $h_{\mu\nu}^{(1)}$,
generally integrated against and contracted into vertices,
\begin{eqnarray}
\lefteqn{h_{\mu\nu}^{(m)}(x) = \int \!\! d^Dx_1 \,
h_{\alpha_1 \beta_1}^{(1)}(x_1) \cdots } \nonumber \\
& & \hspace{2.5cm} \times \int \!\! d^Dx_m \, h_{\alpha_m \beta_m}^{(1)}(x_m) 
\times \mathcal{V}_{\mu\nu}^{~~ \alpha_1 \beta_1 \cdots \alpha_m 
\beta_m}(x;x_1,\dots,x_m) \; . \qquad \label{mthorder}
\end{eqnarray}
It is only the free field $h_{\mu\nu}^{(1)} = \chi_{\mu\nu}^{(1)} +
\psi_{\mu\nu}^{(1)}$ which can be simply decomposed into sub-horizon and 
super-horizon parts. To apportion the higher order contributions we adopt 
the principle of {\it Contagion}, whereby the presence of even a single 
factor of $\psi_{\mu\nu}^{(1)}$ renders the entire term ``super-horizon''.
So substituting $h_{\mu\nu}^{(1)} = \chi_{\mu\nu}^{(1)} + \psi_{\mu\nu}^{(1)}$
in the $m$ free fields of expression (\ref{mthorder}) results in a single 
contribution to $\chi_{\mu\nu}^{(m)}$ from $\chi_{\alpha_1 \beta_1}^{(1)}(x_1) 
\cdots \chi_{\alpha_m \beta_m}^{(1)}(x_m)$ and $2^m -1$ contributions to 
$\psi_{\mu\nu}^{(m)}$.

So the graphs represented in Figures~\ref{Diagram1} and \ref{Diagram2} are 
not quite conventional Feynman diagrams. Because we are taking the expectation 
value of an operator (\ref{Hexpand}) which depends only on the long wavelength 
graviton field $\psi_{\mu\nu}(x)$, the lines emanating from the top point 
$x^{\mu}$ are only the long wavelength ($k < H a(\eta)$) part of the free 
propagator mode sums. By Contagion, all the internal vertices and propagators 
are those of the full theory.

It should not be surprising, and is confirmed by explicit computation
\cite{Miao:2008sp}, that secular enhancements derive entirely from the
purely spatial parts of the free graviton field $\psi_{ij}^{(1)}$. Gravitons
with both indices temporal ($\psi_{00}^{(1)}$), or with mixed time and space
indices ($\psi_{0i}^{(1)}$) make nonzero contributions, but these contributions
do not grow with time. This means we can make a great reduction in the 
expectation value of the order $\kappa^4$ terms in (\ref{Hexpand}),
\begin{eqnarray}
\lefteqn{\hspace{-.2cm} -\frac{\kappa^4}{2} \Bigl\langle \Omega \Bigl\vert 
\psi_{0\rho} \psi^{\rho}_{~ \sigma} \psi^{\sigma}_{~\tau} \psi^{\tau}_{~ 0} 
\!+\! \frac14 (\psi_{0\rho} \psi^{\rho}_{~0} )^2 \!\!+\! \frac12 \psi_{0\rho} 
\psi^{\rho}_{~ \sigma} \psi^{\sigma}_{~ 0} \psi_{00} \!+\! \frac{3}{8} 
\psi_{0\rho} \psi^{\rho}_{~0} \psi_{00}^2 \!+\! \frac{5}{64} \psi_{00}^4 
\Bigr\vert \Omega \Bigr\rangle } \nonumber \\
& & \hspace{1.2cm} \longrightarrow -\frac{\kappa^4}{2} \Bigl\langle 
\Omega \Bigl\vert \psi_{0i}^{(1)}(x) \psi_{0j}^{(1)}(x) \Bigr\vert 
\Omega\Bigr\rangle \Bigl\langle \Omega \Bigl\vert \psi_{ik}^{(1)}(x) 
\psi_{jk}^{(1)}(x) \Bigr\vert \Omega \Bigr\rangle + 
O(\kappa^6) \; . \quad \label{reduction1}
\end{eqnarray}
The order $\kappa^4$ part of (\ref{reduction1}) consists of coincident 
propagators whose evaluation we now discuss.

\subsection{The coincident graviton propagator}

We control ultraviolet divergences with dimensional regularization
in spacetime dimension $D$. Almost all graviton loops on de Sitter have 
been computed using a noncovariant gauge fixing term \cite{Tsamis:1992xa,
Woodard:2004ut},
\begin{equation}
\mathcal{L}_{\rm GF} = -\frac12 a^{D-2} \eta^{\mu\nu} F_{\mu} F_{\nu} 
\quad , \quad
F_{\mu} \equiv \eta^{\rho\sigma} \Bigl[ h_{\mu \rho , \sigma} - \frac12
h_{\rho \sigma , \mu} + (D - 2) a H h^{\mu \rho} \delta^0_{\sigma}
\Bigr] \; . \label{ourgauge}
\end{equation}
The resulting propagator is a sum of three products of a scalar propagator
times a constant tensor factor \cite{Tsamis:1992xa,Woodard:2004ut},
\begin{equation}
i\Bigl[\mbox{}_{\mu\nu} \Delta_{\rho\sigma}\Bigr](x;x') = \sum_{I=A,B,C}
i\Delta_I(x;x') \times \Bigl[\mbox{}_{\mu\nu} T^I_{\rho\sigma}\Bigr] \; .
\end{equation}
The propagators are those for a scalar of masses $m^2_A = 0$, $m^2_B = 
(D-2) H^2$ and $m^2_C = 2(D-3) H^2$. The tensor factors are constructed from 
$\delta^0_{\mu}$ and the spatial part of the Lorentz metric 
$\overline{\eta}_{\mu\nu} \equiv \eta_{\mu\nu} + \delta^0_{\mu} \delta^0_{\nu}$,
\begin{eqnarray}
\Bigl[\mbox{}_{\mu\nu} T^A_{\rho\sigma}\Bigr] & = & 2 \overline{\eta}_{\mu (\rho}
\overline{\eta}_{\sigma ) \nu} - \frac{2}{D \!-\! 3} \, \overline{\eta}_{\mu\nu}
\overline{\eta}_{\rho\sigma} \; , \\
\Bigl[\mbox{}_{\mu\nu} T^B_{\rho\sigma}\Bigr] & = & - 4 \delta^0_{( \mu} 
\overline{\eta}_{\nu ) (\rho} \delta^0_{\sigma )} \; , \\
\Bigl[\mbox{}_{\mu\nu} T^C_{\rho\sigma}\Bigr] & = & \frac{2}{(D \!-\! 2) (D \!-\! 3)}
\Bigl[ (D \!-\! 3) \delta^0_{\mu} \delta^0_{\nu} \!+\! \overline{\eta}_{\mu\nu}
\Bigr] \Bigl[ (D \!-\! 3) \delta^0_{\rho} \delta^0_{\sigma} \!+\! 
\overline{\eta}_{\rho\sigma} \Bigr] \; .
\end{eqnarray}
Note that parenthesized indices are symmetrized.

The full spacetime dependence of all three scalar propagators is known
\cite{Tsamis:1992xa,Woodard:2004ut} but we only require their coincidence
limits. Secular growth comes entirely from the $A$-type propagator \cite{Vilenkin:1982wt,Linde:1982uu,Starobinsky:1982ee},
\begin{equation}
i\Delta_A(x;x) = {\rm Constant} + \frac{H^2}{4 \pi^2} \, \ln(a) \; . 
\label{AVEV}
\end{equation}
At coincidence the $B$-type and $C$-type propagators are actually finite
in dimensional regularization \cite{Woodard:2004ut},
\begin{eqnarray}
i\Delta_B(x;x) & = & -\frac{H^{D-2}}{(4\pi)^{\frac{D}2}} \times 
\frac{\Gamma(D \!-\! 2)}{\Gamma(\frac{D}2)} \longrightarrow 
-\frac{H^2}{16 \pi^2} \; , \label{BVEV} \\
i\Delta_C(x;x) & = & +\frac{H^{D-2}}{(4\pi)^{\frac{D}2}} \times 
\frac{\Gamma(D \!-\! 3)}{\Gamma(\frac{D}2)} \longrightarrow 
+\frac{H^2}{16 \pi^2} \; . \label{CVEV}
\end{eqnarray}

The time dependent part of (\ref{AVEV}) comes entirely from the super-horizon
contributions to the mode sum of the A-type propagator. In contrast, the 
nonzero constants of (\ref{BVEV}-\ref{CVEV}) derive from sub-horizon as well 
as super-horizon modes. However, because the super-horizon mode sum runs from
$k=0$ to $k = H a(\eta) \rightarrow \infty$, we make only a small error (and 
one which falls off with time) in regarding expressions (\ref{BVEV}-\ref{CVEV}) 
as the expectation values of just the super-horizon mode sums. Hence we can
evaluate (\ref{reduction1}) as,
\begin{eqnarray}
\lefteqn{\hspace{-.2cm} -\frac{\kappa^4}{2} \Bigl\langle \Omega \Bigl\vert 
\psi_{0\rho} \psi^{\rho}_{~ \sigma} \psi^{\sigma}_{~\tau} \psi^{\tau}_{~ 0} 
\!+\! \frac14 (\psi_{0\rho} \psi^{\rho}_{~0} )^2 \!\!+\! \frac12 \psi_{0\rho} 
\psi^{\rho}_{~ \sigma} \psi^{\sigma}_{~ 0} \psi_{00} \!+\! \frac{3}{8} 
\psi_{0\rho} \psi^{\rho}_{~0} \psi_{00}^2 \!+\! \frac{5}{64} \psi_{00}^4 
\Bigr\vert \Omega \Bigr\rangle } \nonumber \\
& & \hspace{2cm} \longrightarrow -\frac{\kappa^4}{2} \times i 
\Bigl[\mbox{}_{0i} \Delta_{0j}\Bigr](x;x) \times i \Bigl[\mbox{}_{ik} 
\Delta_{jk}\Bigr](x;x) + O(\kappa^6) \; , \qquad \\
& & \hspace{1cm} = -\frac{\kappa^4}{2} \times -\delta_{ij} i\Delta_B(x;x) 
\times \Bigl(D - \frac{2}{D \!-\! 3}\Bigr) \delta_{ij} i\Delta_A(x;x) +
O(\kappa^6) \; , \qquad \\
& & \hspace{2cm} \longrightarrow -\frac{3 (\kappa H)^4}{64 \pi^4} \, 
\ln(a) + O(\kappa^6) \; . \label{reduction2}
\end{eqnarray}

Although the initial expansion (\ref{Hexpand}) of $\mathbf{H}/H$ in powers
of the graviton field is the same in all gauges, the evaluation of the
expectation value of individual terms in this expansion, such as 
(\ref{reduction1}), is of course dependent upon the gauge. We have chosen
to work in the covariant gauge (\ref{ourgauge}), which is the simplest to
use. However, it is worth describing how the same effect would appear in
the much more complicated formalism associated with a physical gauge such
as $\partial_i \psi_{ij} = 0 = \psi_{ii}$. In that gauge the 
spatial-transverse-traceless components of the graviton field $\psi^{TT}_{ij}$
are invariant under linearized gauge transformations, although they are not 
fully invariant. The other components of the metric, including $\psi_{00}$ 
and $\psi_{0i}$, are also not zero. They would be expressed as series expansions
in powers of $\psi_{ij}^{TT}$, staring at order $\kappa \psi^2$, by 
perturbatively solving the constraint equations. So the same order $\kappa^4$ 
effect that we obtained in expression (\ref{reduction2}) might derive from the 
expectation value of the terms $\kappa^2 [\frac38 \psi_{00}^2 - \frac12 \psi_{0i} 
\psi_{0i}]$ from expression (\ref{Hexpand}). Note that the gauge-fixed and 
constrained Lagrangian of a physical gauge is not local, and can of course only 
be expressed to some finite order because exact solutions for the constraint 
equations are not known. That is what makes this formalism so terrifically 
difficult to use. In fact all graviton loop computations on de Sitter background 
have been performed using covariant gauges for which the Lagrangian is local and 
the graviton propagator includes both constrained and physical components. 

\subsection{IR cleansed Hubble parameter \& field equation}

Recall that expression (\ref{Hexpand}) for $\mathbf{H}(x)/H$ contains 
terms with $N$ factors of $\psi_{\mu\nu}$ for $N = 0, 1, 2, \dots$
The expectation value has constant contributions at order $\kappa^2$ from 
the $N=1$ \cite{Tsamis:2005je} and $N=2$ terms, which can be absorbed into a 
renormalization of the cosmological constant. Expression (\ref{reduction2}) 
gives the secular contribution from the $N = 4$ term. If we assume there 
are no secular contributions at order $\kappa^4$ from the $N =1$, $N=2$ and 
$N=3$ terms (more on this later) then the expectation value of 
$\mathbf{H}(x)$ is,
\begin{equation}
\Bigl\langle \Omega \Bigl\vert \mathbf{H}(x) \Bigr\vert \Omega \Bigr\rangle
\longrightarrow H \Biggl\{ 1 - \frac{3 (\kappa H)^4}{64 \pi^4} \, \ln(a) 
+ O(\kappa^6) \Biggr\} \; . \label{HVEV}
\end{equation}
It is interesting to note that secular slowing is predicted to occur at
the same order, and with the same time dependence, when one does not 
excise the super-horizon modes but rather includes their contribution to
the vacuum energy \cite{Tsamis:2011ep}.

It is worth digressing at this point to note that our result (\ref{HVEV})
applies even to pure quantum gravity, with a positive cosmological constant,
released in Bunch-Davies vacuum. One consequence is that gravitons do not
possess a fully de Sitter invariant vacuum state, just like the massless, 
minimally coupled scalar \cite{Allen:1987tz}, whose plane wave mode functions
are identical to those of dynamical gravitons \cite{Lifshitz:1945du}. There 
has been a long and confusing debate about this \cite{Higuchi:2011vw,Miao:2011ng,
Morrison:2013rqa,Miao:2013isa,Frob:2014fqa,Woodard:2015kqa}. All agree that 
the graviton mode functions approach a constant at late times (that is what 
causes the tensor power spectrum) and that this freezing-in endows the 
completely gauge fixed graviton propagator with a de Sitter-breaking time 
dependence which takes the form of a linearized gauge transformation. The 
debate concerns whether or not this time dependence can have physical 
consequences analogous to those of the constant gauge field in the famous 
Aharonov-Bohm effect \cite{Aharonov:1959fk}. Our attitude is to decide the
matter by computation, using the propagator described in section 3.2 which all
sides accept as valid. Our result (\ref{HVEV}) does support the view that de 
Sitter breaking is real, although this conclusion needs to be confirmed by a 
complete, two loop computation of an invariant measure of the local expansion 
rate \cite{Miao:2017vly}.

A final comment concerns the magnitude and universality of the effect. Even 
during primordial inflation, the dimensionless loop counting parameter is 
minuscule, $\kappa^2 H^2 < 10^{-10}$. However, the factor of $\ln(a)$ grows
with time so that the effect must eventually become nonperturbatively strong. 
Expression (\ref{HVEV}) was derived using perturbation theory, hence it is 
valid so long as $\kappa^2 H^2 \ln(a)$ is small. This means that it applies 
to the early stages of inflation for any vacuum energy of a few orders of 
magnitude below the Planck mass, all the way down to zero.

The original metric $g_{\mu\nu}(x)$ obeys the exact Heisenberg field equation
$R(x) = D \Lambda$, where the cosmological constant $\Lambda$ is $(D-1) H^2$
plus renormalization counterterms. That fact has been invoked to claim that
there can be no back-reaction \cite{Garriga:2007zk,Tsamis:2008zz}. However,
it could be argued that one must instead re-organize the operators of the Ricci 
scalar so as to extract quantum corrections to the vacuum energy, the same way 
one does for the generator $L_0$ in Virasoro algebra of free string theory
\cite{Goddard:1972iy}. In that case back-reaction would derive from integrating 
the vertices of loop corrections back to the initial value surface, over the 
larger and larger past light-cones which open as the observation point $x^{\mu}$ 
occurs later and later after the initial value surface. Although the process is 
completely causal in spacetime, it does involve contributions from the 
super-horizon modes $\psi_{\mu\nu}$ of the graviton field, which disturbs those 
who believe in the Transformation Ansatz. We therefore examine the field 
equation obeyed by the local observer's metric (\ref{localmetric}) which is 
free of super-horizon modes.

It is useful to extract the local observer's scale factor (\ref{localscale}) 
from his metric (\ref{localmetric}), 
\begin{equation}
\mathbf{g}_{\mu\nu}(x) \equiv \mathbf{a}^2(x) \times 
\widehat{\mathbf{g}}_{\mu\nu}(x) \qquad , \qquad \mathbf{a}(x) \equiv 
-\frac1{ \mathbf{H}(x) \, \eta} \; .
\end{equation}
The local observer's Ricci tensor follows from a conformal transformation,
\begin{eqnarray}
\lefteqn{\mathbf{R}_{\mu\nu} = \widehat{\mathbf{R}}_{\mu\nu} - (D \!-\! 2) \Bigl(
\frac{\mathbf{a}_{,\mu}}{\mathbf{a}} \Bigr)_{; \nu} - 
\widehat{\mathbf{g}}_{\mu\nu} \widehat{\mathbf{g}}^{\rho\sigma} \Bigl(
\frac{\mathbf{a}_{,\rho}}{\mathbf{a}} \Bigr)_{; \sigma} } \nonumber \\
& & \hspace{4.5cm} + (D \!-\! 2) \frac{\mathbf{a}_{,\mu}}{\mathbf{a}} 
\frac{\mathbf{a}_{, \nu}}{\mathbf{a}} - (D \!-\! 2) \widehat{\mathbf{g}}_{\mu\nu} \widehat{\mathbf{g}}^{\rho\sigma} \frac{\mathbf{a}_{, \rho}}{\mathbf{a}} 
\frac{\mathbf{a}_{, \sigma}}{\mathbf{a}} \; , \qquad \label{localRicci}
\end{eqnarray}
where a comma denotes ordinary differentiation and a semicolon indicates
covariant differentiation with the affine connection of 
$\widehat{\mathbf{g}}_{\mu\nu}$. If we ignore the small spacetime variation 
of $\mathbf{H}(x)$ then derivatives of the scale factor are,
\begin{eqnarray}
\frac{\mathbf{a}_{,\mu}}{\mathbf{a}} & \longrightarrow & 
-\frac{\delta^0_{~\mu}}{\eta} = \mathbf{H} \mathbf{a} \delta^0_{~\mu} \; , \\
\Bigl( \frac{\mathbf{a}_{,\mu}}{\mathbf{a}} \Bigr)_{; \nu} & \longrightarrow &
\mathbf{H}^2 \mathbf{a}^2 \delta^0_{~\mu} \delta^0_{~\nu} - 
\widehat{\mathbf{\Gamma}}^0_{~\nu\mu} \mathbf{H} \mathbf{a} \longrightarrow
\mathbf{H}^2 \mathbf{a}^2 \delta^0_{~\mu} \delta^0_{~\nu} \; ,
\end{eqnarray}
where the final simplification comes from retaining only terms with the largest
number of scale factors. With the same approximations we have,
\begin{equation}
\mathbf{R}_{\mu\nu} \longrightarrow -(D \!-\! 1) \widehat{\mathbf{g}}_{\mu\nu} 
\widehat{\mathbf{g}}^{00} \mathbf{H}^2 \mathbf{a}^2 = -(D \!-\! 1) \mathbf{H}^2
\widehat{\mathbf{g}}^{00} \times \mathbf{g}_{\mu\nu} \; . \label{localeqn}
\end{equation}
This is the Einstein equation with a time-dependent cosmological constant,
\begin{equation}
\mathbf{\Lambda}(x) = (D \!-\! 1) \mathbf{H}^2(x) \times 
-\widehat{\mathbf{g}}^{00}(x) \; . \label{localLambda}
\end{equation}
So invoking the Transformation Ansatz does not avoid the reality of
back-reaction.

\section{Epilogue}

The conventional way of thinking about inflationary back-reaction is that
inflation continually rips long wavelength gravitons out of the vacuum 
and the self-gravitation between them slows the expansion rate by an 
ever-increasing amount as more and more of these gravitons come into 
causal contact \cite{Tsamis:1996qq,Tsamis:2011ep}. One objection is that
the Heisenberg field equations imply the Ricci scalar is constant
$R = D \Lambda$ \cite{Garriga:2007zk,Tsamis:2008zz}. However, the Ricci
scalar --- and any other nonlinear field operator --- diverges when 
acting on physical states, hence one should order it so as to extract the
vacuum energy, the same way one does for $L_0$ in the Virasoro algebra of
free string theory \cite{Goddard:1972iy}. When that is done, back-reaction
manifests as diagrams which make secular contributions to the vacuum energy.

Although the diagrams which contribute to secular back-reaction are 
completely causal in spacetime, they do derive from Fourier components of 
the graviton field operator whose physical wavelengths (in the background 
geometry) exceed the instantaneous Hubble radius. This occasions intense 
scepticism \cite{Urakawa:2009my,Urakawa:2009gb}, and has led to assertions 
that super-horizon modes are not accessible to a local observer, but would 
instead be subsumed into a transformation of his coordinate system \cite{Giddings:2010nc,Urakawa:2010it,Tanaka:2011aj,Giddings:2011zd}. The
argument then runs that secular back-reaction is impossible because the
local observer's metric does not even possess any of the super-horizon 
modes which might cause it.

We refer to this belief as the {\it Transformation Ansatz} and we showed 
in section 2 that the local observer's metric (\ref{localmetric}) it 
implies is properly a nonlocal field redefinition of the original metric.
Secular back-reaction still occurs because the transformation which absorbs 
the super-horizon gravitons changes the conformal time coordinate upon which 
the scale factor depends. In section 3 we constructed the expansion rate 
(\ref{localH}) and field equation (\ref{localeqn}) which would be perceived 
by a local observer. Both of these quantities are operators so one can only 
discuss their statistical properties. With one assumption we were able to 
evaluate their expectation values (\ref{HVEV}), which show secular slowing 
at exactly the same order and with the same time dependence that is predicted 
in the conventional picture \cite{Tsamis:2011ep}. We conclude that the 
Transformation Ansatz does not preclude but rather confirms secular 
back-reaction.

The assumption we made to derive (\ref{HVEV}) is a large one: that 
there are no secular contributions at order $\kappa^4$ from the 1-point, 
2-point and 3-point diagrams of Figure~\ref{Diagram1}. We doubt that this 
can be correct. However, even if the other diagrams change the result 
(\ref{HVEV}), the fact remains that the local observer's expansion rate 
$\mathbf{H}(x)$ is both dynamical and time-dependent. Note that 
$\mathbf{H}(x) = H/N(x)$, where $N(x)$ is the ADM lapse of the
super-horizon gravitons, so one interpretation of secular back-reaction is 
that the continual freezing-in of modes gradually increases the time scale.

It seems to us that adopting the Transformation Ansatz is problematic
because it denies the locality of interactions, and because it makes the
number of degrees of freedom depend upon the background geometry and vary
with time. However, we have been careful not to pronounce on the validity
of the Transformation Ansatz; our point is merely that adopting it leads
to secular back-reaction of the same sort that is predicted to occur when
the super-horizon modes are retained. There seem to be complementary 
pictures:
\begin{itemize}
\item{One can either employ the full metric --- including super-horizon 
modes --- and then one sees a vacuum energy whose time dependence derives 
from more and more modes coming into interaction; or}
\item{One can excise the super-horizon modes, and then one sees a 
time-dependent expansion rate from a gradual increase in the lapse which 
sets the scale of time.}
\end{itemize}

Although we are dubious as to the validity of the Transformation Ansatz 
there is no doubt that local observers couple only weakly to individual
super-horizon modes. Hence, it may not be a bad approximation to assume 
that local observers perceive the geometry of the cleansed metric 
(\ref{localmetric}). Perhaps there is a sort of spacetime uncertainty 
principle at work: One could indeed infer the curvature by measuring the 
geodesic separation between freely falling observers, but resolving the 
contributions from modes of longer and longer wavelengths requires longer 
and longer times. So the result obtained within a Hubble time is the 
curvature of the infrared-cleansed metric (\ref{localmetric}).

Finally, we would like to suggest that the increase in the gravitational
lapse associated with the horizon crossing of a graviton mode can be viewed
as a cosmological analogue of the famous gravitational ``memory effect''
\cite{Zeldovich:1974,Braginsky:1986ia,Ludvigsen:1989kg,Christodoulou:1991cr}.
Recall that the memory effect is a permanent shift in the geodesic separation
between freely falling observers who experience the passage of a gravitational
wave. The curvature is zero before and after the wave, yet the shift in their
locations is real. {\it Note that the small curvature associated with 
super-horizon gravitons is one of the chief arguments against them having any
effect during inflation, and this very same argument could be invoked to deny 
the reality of the memory effect due to gravitational waves.} That argument
was wrong in flat space background and there is no reason to take it any more 
seriously during inflation. 

The analogy with the memory effect is worth pursuing a little further. It has 
been shown \cite{Strominger:2014pwa} that the positional offset induced by the 
passage of a gravitational wave can be expressed as the action of a BMS 
transformation \cite{Bondi:1962px,Sachs:1962wk}, a class of diffeomorphisms 
which does not go to zero at spatial infinity. The cosmological analogue of
these transformations has been constructed \cite{Hinterbichler:2013dpa} and
their action has been shown to add a super-horizon graviton 
\cite{Ferreira:2016hee}. The effect which interests us is not this linear
one but rather a higher order part of what is the same transformation, so
it is good to know that the full nonlinear extension exists. We should
also note that the ability to absorb super-horizon gravitons using these
infinite range diffeomorphisms was previously invoked by those who dispute 
the reality of secular graviton effects \cite{Higuchi:2011vw}. It will be 
seen that the very same argument could be used to deny the reality of 
gravitational memory and is therefore falsified.

\vskip 1cm

\centerline{\bf Acknowledgements}

This work was partially supported by the European Union's Seventh
Framework Programme (FP7-REGPOT-2012-2013-1) under grant agreement
number 316165; by the European Union's Horizon 2020 Programme
under grant agreement 669288-SM-GRAV-ERC-2014-ADG; by NSF grant 
PHY-1506513; and by the Institute for Fundamental Theory at the 
University of Florida.

\end{document}